\documentstyle[epsfig]{aipproc}

\begin{document}
\title{Hydrogen-Bonded Liquids: Effects of Correlations of Orientational
       Degrees of Freedom }

\author{Giancarlo Franzese$^*$, Masako Yamada$^*$${\dagger}$\,
	 and H.~Eugene Stanley$^*$.\\}

\address{$^*$Center for Polymer Studies and Department
of Physics, and $^{\dagger}$Center for Computational ~Science Boston
University, Boston MA 02215, USA.\\ }

\maketitle

\begin{abstract}

We improve a lattice model of water introduced by
 Sastry, Debenedetti, Sciortino, and
Stanley to give insight on experimental
 thermodynamic anomalies in supercooled phase,
taking into account  the
correlations between intra-molecular orientational degrees of freedom.
The original Sastry et al. model including energetic, entropic
and volumic effect of the orientation-dependent hydrogen bonds (HBs),
captures qualitatively the experimental water behavior, but it ignores
the geometrical correlation between HBs.
Our mean-field calculation shows that adding these correlations gives a
more water-like phase diagram than previously shown, with the appearance
of a solid phase and first-order liquid-solid  and gas-solid phase
transitions. 
Further investigation is necessary to be able to use this model
to characterize the thermodynamic properties of the supercooled region.

\end{abstract}

\section*{Introduction}

Water has a temperature density maximum at 4~$^o$C at ambient pressure,
above the melting line. This is perhaps the most well-known anomaly of the
many observed in liquid water. In the metastable, supercooled region of
water, the anomalies are especially pronounced. The absolute magnitude of
thermodynamic response functions such as isothermal
compressibility~\cite{Speedy76}, thermal expansion
coefficient~\cite{Hare86}, and isobaric heat capacity~\cite{Angell82} are
all known to increase dramatically as the temperature is lowered. The
quest for an explanation has led to extensive experimental, numerical, and
theoretical investigations of the properties of supercooled
water~\cite{AngellChapter,BellissentBook,Bellissent95,Mishima98,Smith99,Shiratani98,DebenedettiBook,Ponyatovsky95,Roberts96a,Roberts96b,Reza98}
However, in spite of many studies, a
final answer has yet to be reached.

\begin{figure}
\centerline{ \epsfig{file=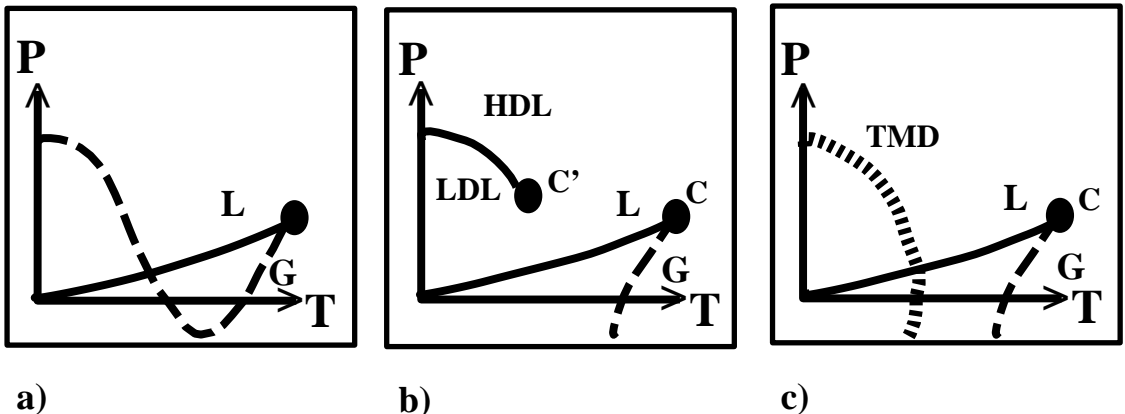} }
\caption{Three interpretations of the supercooled water phase diagram: a)  
stability limit conjecture, b) two critical-points model, c)
singularity-free scenario. A dashed line denotes a spinodal, a bold line
denotes a transition line, and a dotted line denotes a TMD line. The
solid-liquid transition line is not shown;
$C$ is the known
critical point; $C'$ is the hypothetical second critical point.}
\label{fig1} 
\end{figure}

Three major theories have emerged from existing evidence. They are known
as the stability-limit conjecture~\cite{Speedy82}, the two critical-point
model\cite{Poole92}, and the singularity-free
scenario~\cite{Stanley80,Sastry96,Rebelo98} (Fig.~\ref{fig1}). 
$i)$ The stability-limit conjecture assumes
that the superheated liquid-to-gas spinodal and the supercooled
liquid-to-solid spinodal in the pressure-temperature ($P,T$) 
phase diagram are connected at
negative $P$. 
$ii)$ The two critical-point scenario does not assume such a
{\em retracing spinodal}. Instead, it attributes the anomalies of water to
a second critical-point in the supercooled region of the phase diagram.
This critical-point comes at the end of a phase-transition line that
separates two metastable liquid phases, a low-density liquid (LDL), and a
high-density liquid (HDL). 
$iii)$ The singularity-free scenario does not
attribute the anomalies in water to a transition in the supercooled
region or a retracing spinodal, but to a line of temperatures of maximum
density (TMD) in the stable liquid phase and in the super-heated
 liquid phase with positive
$(\partial P/\partial T)_{TMD}$ (slope)
in the $P,T$ plane at high $P$ and negative slope at lower $P$
({\em retracing} TMD). The aforementioned response functions are not considered to
diverge, but merely to have maxima.
Note that even in the two critical points scenario the TMD is retracing.
There are theoretical calculations that show that the apparent gap between
these different scenarios can be reconciled without
inconsistencies~\cite{Poole94,Borick95,Roberts96b}. 
Thus, the differences between
the different scenarios may be much more subtle than expected.
Authoritative experiments have not been forthcoming due to the difficulty
of probing metastable water without encountering inevitable
crystallization \cite{Mishima98}.

It is worth noting that these so-called anomalies are not unique to water.
Although water exhibits anomalous properties relative to the liquid norm,
namely argon-like liquids, its unusual characteristics are shared by other
fluids such as $SiO_2$\cite{Grims84} and $GeO_2$\cite{Smith95}. What
differentiates these anomalous liquids from normal liquids is that they
form bonded networks that are dependent on the relative orientation of the
molecules.  

The effect of orientation-dependent, hydrogen-bonded networks on critical
behavior has already been recognized. However, the effect of the
\emph{intra-}molecular correlation between the orientational degrees of
freedom has not be explored thoroughly. For example, an open problem is
how the intensity of orientational correlations can change the phase
diagram. This is where our focus lies.

\section*{The Model}

Our Hamiltonian can be expressed as:
\begin{equation}
\label{hamil}
{\cal H}=-\epsilon\sum_{<i,j>}{n_in_j}-\mu \sum_i n_i
    -\it{J}\sum_{<i,j>}{n_in_j\delta_{\sigma_{ij}\sigma_{ji}}}
    -\it{J_\sigma}\sum_in_i\sum_{(k,l)_i}{\delta_{\sigma_{ik}\sigma_{il}}} \ .
\end{equation}
The first three terms constitute the original Hamiltonian \cite{Sastry96,Rebelo98}.  
The first two terms in the Hamiltonian are the standard lattice gas
Hamiltonian  where
$n_i = 0, 1$ indicates whether site $i$ is occupied by a molecule,
$\epsilon$ is the attractive energy between molecules, $\mu$ is the
chemical potential and the symbol
$\sum_{<i,j>}$ means that the sum is on all nearest neighbor (NN)
lattice sites. 

The third term in Eq.(\ref{hamil}) describes the interaction of
orientational degrees of freedom (arms) 
\emph{between} neighboring molecules, and accounts for hydrogen
bonding within the system. It is introduced in the following way: Each
molecule has a maximum of $\gamma$ neighbors; Therefore, it has $\gamma$
 arms that can potentially form HBs with the respective
arms of its neighbors. Each of the arms can assume $q$ orientational
states, so the total number of orientational states for a molecule 
 is $q^\gamma$. The Potts variable $\sigma_{ij} = 1,2,\ldots,
q$ \cite{Potts} specifies the orientational state of the arm of molecule $i$ that
points toward nearest-neighbor molecule $j$. When two neighboring sites
are occupied ($n_i=n_j=1$) and the orientational states of the two
reciprocally pointing arms match ($\sigma_{ij}=\sigma_{ji}$), a HB
forms ($\delta_{\sigma_{ij}\sigma{ji}}=1$ where $\delta_{ab}=1$ if
$a=b$, otherwise it is $\delta_{ab}=0$). Otherwise is 
$n_in_j\delta_{\sigma_{ij}\sigma_{ji}}=0$.
This leads to an energy gain $J$ per HB. This third term is
also summed over all NN sites.

The fourth term in Eq.(\ref{hamil}) comes from our generalization, and takes
into consideration the interaction of orientational degrees of freedom
\emph{within} each molecule. For an occupied site $n_i=1$, there is an
energy gain ${\it J_\sigma}$ if the orientations of two arms, $k$ and $l$,
are not independent of each other (for the sake of simplicity, they have
to be in the same state in this model). If ${\it J_\sigma}=0$ we recover
the original Sastry et al. Hamiltonian \cite{Sastry96,Rebelo98}. This term is summed
over all sites, and over all intra-molecular permutations of arms within
each molecule.

As in the Sastry et al. model, in our generalization the creation of a HB 
not only contributes to an energy factor, but also leads to an
increase in volume:
\begin{equation}
\label{vol}
V=V_o + dV N_{HB}
\end{equation}
where $V_o$ denotes the volume of the system without HBs, $dV$
denotes the increase in volume per HB, and $N_{HB}$ denotes the
total number of HBs in the system. The original model takes
into account main ingredients: $i)$ The increase of volume upon HB
formation (Eq.\ref{vol}); $ii)$ The decrease of entropy upon HB formation
due to the decrease of orientational states available to
hydrogen bonded molecules. To these two features we add a third
ingredient that is: $iii)$ The correlation between the
orientations of different arms of the same molecules.

\section*{The Mean-Field Calculation}


From the above explanations, we can see that there are two order parameters in
this system, a lattice gas order parameter $m\in[-1,1]$, and an orientational order
parameter $s\in[0,1]$. 
The order parameter $m$ indicates the proportion of
occupied sites and is related to the fraction of occupied sites
(number of molecules
normalized to the total number of sites $N$) $n=\sum_i n_i/N$
but it does not give the true density of the system
$\rho=nN/V$ where $V$ is given by Eq.(\ref{vol}), affected by
the number of HBs. The orientational order parameter $s$ is
required to find the true density of the system.

The orientational order parameter $s$ is a measure of how much the
preferred orientational state dominates the system. 
When $s=0$, all of the arms of occupied sites assume a random
orientational state from among the $q$ possible states. No orientational
state is favored and network-formation is not encouraged. When $s=1$, all
of the arms assume the same orientational state. Thus, hydrogen-bonded
network formation is heavily favored, provided sufficient number density.
This order parameter is analogous to the magnetization of an Ising system,
where there is no preferred orientational state initially, but where one
state is spontaneously ``chosen'' due to random fluctuations (symmetry breaking).

In mean field (MF) approximation the fraction of occupied sites $n$ is
supposed to be a linear function of the lattice gas order parameter
$m$ as $n=(1+m)/2$ and, analogously, the fraction of Potts
variables in a given state (supposing a symmetry breaking) is
$y=[1+(q-1)s]/q$ as function of the Potts MF order parameter
$s$. 

To find the equilibrium points we consider the MF molar Gibbs free energy
per site $g=u-Ts+Pv\equiv \mu$ where 
\begin{equation}
u=-\frac{\gamma}{2}\left[\epsilon n+\left(Jn+\frac{\gamma-1}{2}\gamma 
J_\sigma\right)\left(y^2+\frac{(1-y)^2}{q-1}\right)\right]
\label{u}
\end{equation}
is the molar energy per site,
\begin{equation}
s=-k_B\left[\ln n+\frac{1-n}{n}\ln(1-n)+\gamma\left(y\ln
y+(1-y)\ln\frac{1-y}{q-1}\right)\right]
\label{s}
\end{equation}
is the molar entropy per site ($k_B$ is the Boltzmann constant)
calculated considering that any molecule can have $q^\gamma$ states, 
\begin{equation}
v=\frac{v_0}{n}+\frac{\gamma}{2}nv_{HB}\left(y^2+\frac{(1-y)^2}{q-1}\right)
\label{v}
\end{equation}
is the molar volume per site ($v\equiv1/\rho$) with $v_0=V_0/N$ and
$v_{HB}=dV/N$. 

Using temperature and pressure as input variables,
$g$ is minimized with respect to the order parameters. The resulting
values for $m$ and $s$ are directly related to the number density and
hydrogen-bonding probability of that state point. Together, these can
provide the true density, which is needed to map out the phase diagram.

\begin{figure}[ht]
\begin{minipage}{3in}
\epsfig{file=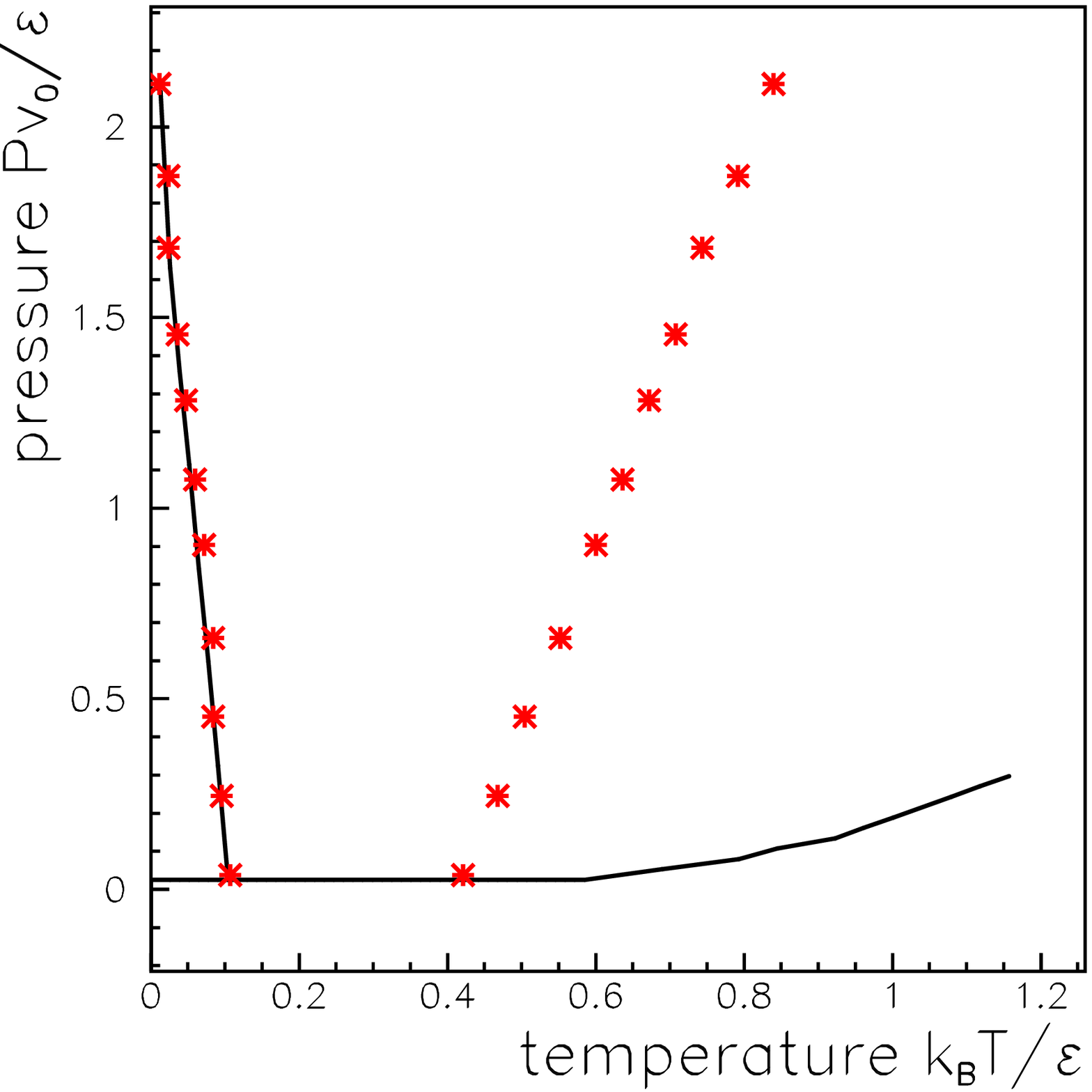,height=3in,width=3in}
a)
\end{minipage}\hspace*{0.05in}
\begin{minipage}{3in}
\epsfig{file=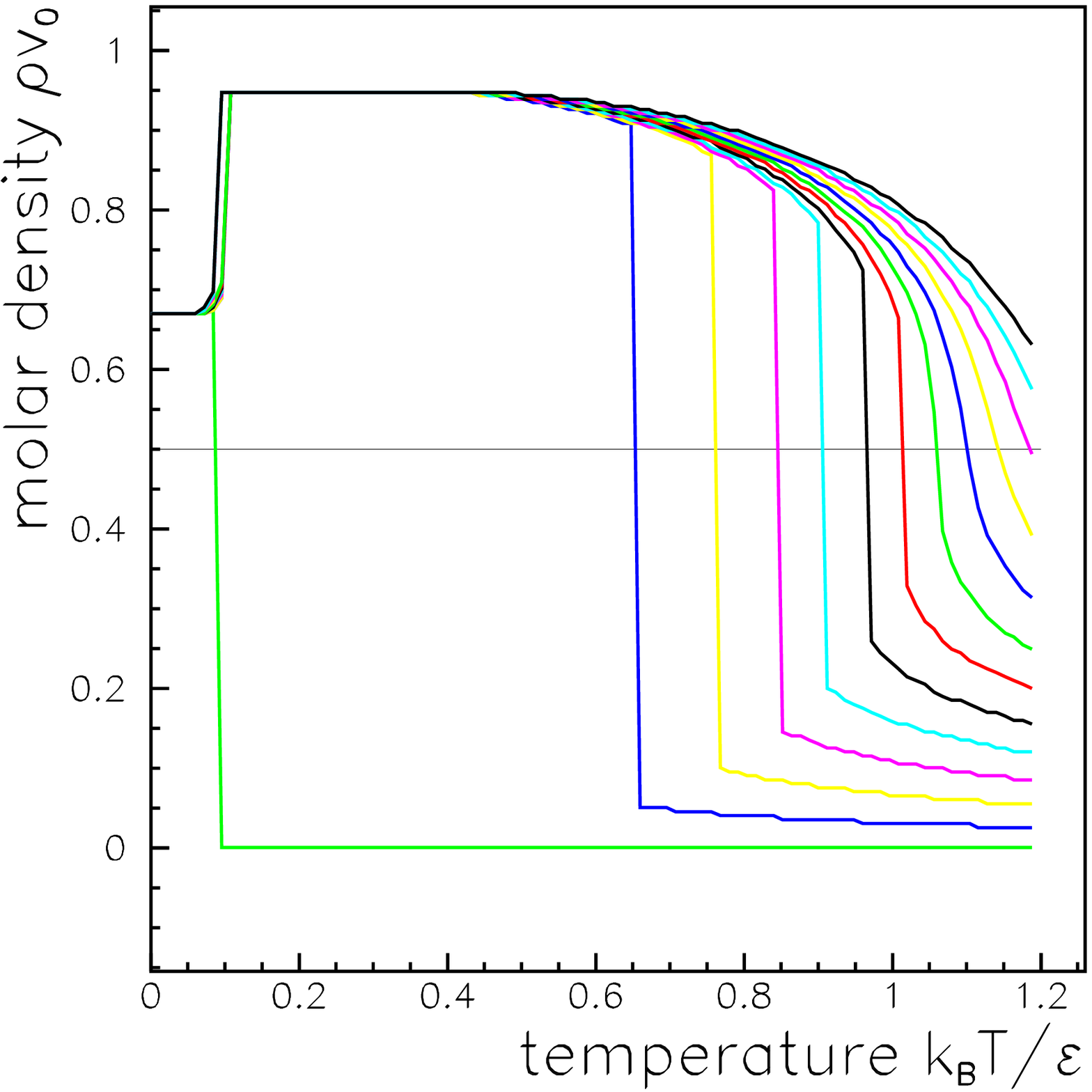,height=3in,width=3in} 
b)
\end{minipage}
\caption{
MF phase diagram for the model in 
Eqs.(\protect\ref{hamil},\protect\ref{vol}) with 
$J/\epsilon=0.5$, $X/\epsilon=0.01$, $v_{HB}/v_0=0.25$, $\gamma=4$,
$q=10$.  a) The $P,T$  projection: 
Lines represent phase transitions with negatively sloped solid-liquid
transition and liquid-gas transition ending
in a critical point; Stars represent the  limits of the TMD
region.
 b) The $\rho,T$ projection for isobars ranging from
$Pv_0/\epsilon=-.0007$ (bottom) to $Pv_0/\epsilon= 
0.36$ (up) with increment of $\Delta Pv_0/\epsilon=0.03$.
Note that for $Pv_0/\epsilon=-.0007$ there is a solid-gas transition, that is
indistinguishable in a).
}
\label{fig2} 
\end{figure}

Previous two-dimensional mean-field calculations executed by Sastry et al.  
using Eq.(\ref{hamil}) without the fourth term, have shown that an
orientationally-dependent, network-bonded liquid without correlations
between the orientational degrees of freedom exhibits phase behavior as
described by the singularity-free scenario \cite{Sastry96,Rebelo98}.
(Fig.\ref{fig1}c). 
When the fourth term of Eq.(\ref{hamil}) is included
an additional water-like features appear in the phase
diagram as shown in Fig.\ref{fig2}.
Namely, the MF calculation recovers a new full-bonded ($s=1$) phase 
that represents  the solid phase. In particular, we find a positively
sloped first-order solid-gas transition line and a negatively sloped 
first-order solid-liquid transition line ending in a 
tricritical point. 
Furthermore 
it is possible to recover the same behavior for
the liquid-gas spinodal line (not shown in Fig.\ref{fig2}).
But, in the limit of this approach, is
not possible to see for each pressure a single point of maximum density
(the TMD line),
but only a range of temperatures in which the density (molar volume) has
a flat maximum (minimum). For each pressure the lower temperature of
this range is indistinguishable on the scale of Fig.\ref{fig2} to the
solid-liquid transition line, while the higher temperature is shown in
Fig.\ref{fig2} and has a positive slope. This result is not very
different from the MF result found for the Sastry et al. Hamiltonian with
the assumption of no correlation between orientational degrees of
freedom, shown in Fig.5 of Ref.\cite{Rebelo98}, where 
the higher is the pressure the flatter is the 
minimum in the molar volume. The only difference is that in our case a
flat minimum is hardly distinguished by a range of minima.
Therefore this model in this MF approach leads to the 
same  singularity-free
scenario for the density anomalies of the water already found in the
original model of Sastry et al. \cite{Sastry96,Rebelo98}.

It is to be noted here that in our approach we are ignoring the
fluctuations and supposing that the Potts (orientational) 
symmetry breaking always occurs, therefore we cannot have a new
transition line ending in a new critical point. The inclusion of
fluctuations could be the essential feature to recover the
second critical point in the metastable supercooled liquid phase.
Furthermore at negative pressure and $T=0$ a spurious state appears, due
to the strong assumption (the Potts symmetry breaking) that we made and
to the fact that upon stretching the system can 
minimize the Gibbs free energy forming HBs in the gas phase.

\section*{Conclusions and Open Issues}

By taking a simple model of water that leads to the singularity-free
scenario \cite{Sastry96,Rebelo98} as interpretation of the anomalous
water behavior  and adding a
term that correlates the intra-molecular orientational degrees of freedom,
we recover
the solid phase,  the liquid-solid and the gas-solid transition lines.
Furthermore the TMD ``region'' is above the melting line.
All those are characteristic features of water
not included in the seminal Sastry et al. model \cite{Sastry96,Rebelo98}.
This can be considered a more realistic
extension of the previously mapped singularity-free phase diagram, which
has no such transitions.
However, these are stable transitions which
by themselves cannot verify or negate the existence of singularities in the
metastable region, but they make the phase diagram more water-like. 

So far our investigation is done only adding
correlations between the orientational degrees of freedom in the limit of no
fluctuations. 
In order to understand the critical phenomena of water in  
the supercooled region, thermal fluctuations cannot be ignored.
An inherent shortcoming of MF calculations is that such
fluctuations are blotted out by strong approximations. 
One way of including such fluctuations into the analysis is to consider them
into the MF  calculations in a perturbative manner,
for instance by taking into account the local fluctuations of density
\cite{Borick95}. Another way is to conduct numerical simulations
\cite{Roberts96a}. 
We will now pursue these two new paths to see what effect
fluctuations have on the phase behavior of this system, particularly in
the metastable region.

\section*{Acknowledgments}

We would like to thank A.~Scala for many enlightening discussions. G.F. is
partially supported by CNR grant N.203.15.8/204.4607 (Italy), M.Y. is supported by a
National Science Foundation traineeship administered by the Center for
Computational Science at Boston University, and the Center for Polymer
Studies is supported by the National Science Foundation.

\end{document}